\documentclass[sigconf]{acmart} 

\usepackage{amsmath,amsfonts}
\usepackage{algorithmic}
\usepackage{graphicx}
\usepackage{textcomp}
\usepackage{xcolor}
\usepackage{url}
\usepackage{booktabs}
\usepackage{enumitem}
\usepackage[para]{footmisc}
\setcopyright{acmcopyright}
\copyrightyear{2024}
\acmYear{2024}
\setcopyright{rightsretained}
\acmConference[EASE 2024]{28th International Conference on Evaluation and Assessment in Software Engineering}{June 18--21, 2024}{Salerno, Italy} \acmBooktitle{28th International Conference on Evaluation and Assessment in Software Engineering (EASE 2024), June 18--21, 2024, Salerno, Italy}\acmDOI{10.1145/3661167.3661212}

\acmISBN{979-8-4007-1701-7/24/06}

\begin{document}

\title[SCM2 Reference Architecture]{An Empirically Grounded Reference Architecture for \\Software Supply Chain Metadata Management
}

\author{Nguyen Khoi Tran}
\affiliation{%
  \institution{\textit{CREST} \\
 \textit{The University of Adelaide}}
  \city{Adelaide}
  \country{Australia}}
\email{nguyen.tran@adelaide.edu.au}

\author{Samodha Pallewatta}
\affiliation{%
  \institution{\textit{CREST} \\
 \textit{The University of Adelaide}\\}
  \city{Adelaide}
  \country{Australia}}
\email{samodha.pallewatta@adelaide.edu.au}

\author{M. Ali Babar}
\affiliation{%
  \institution{\textit{CREST, The University of Adelaide} \\
\textit{Cyber Security Cooperative Research Centre}\\}
  \city{Adelaide}
  \country{Australia}}
\email{ali.babar@adelaide.edu.au}

\begin{abstract}

With the rapid rise in Software Supply Chain (SSC) attacks,  organisations need thorough and trustworthy visibility over the entire SSC of their software inventory to detect risks early and identify compromised assets rapidly in the event of an SSC attack. One way to achieve such visibility is through SSC metadata, machine-readable and authenticated documents describing an artefact's lifecycle. Adopting SSC metadata requires organisations to procure or develop a Software Supply Chain Metadata Management system (SCM2), a suite of software tools for performing life cycle activities of SSC metadata documents such as creation, signing, distribution, and consumption. Selecting or developing an SCM2 is challenging due to the lack of a comprehensive domain model and architectural blueprint to aid practitioners in navigating the vast design space of SSC metadata terminologies, frameworks, and solutions. 
This paper addresses the above-mentioned challenge by presenting an empirically grounded Reference Architecture (RA) comprising of a domain model and an architectural blueprint for SCM2 systems. Our proposed RA is constructed systematically on an empirical foundation built with industry-driven and peer-reviewed SSC security frameworks. Our theoretical evaluation, which consists of an architectural mapping of five prominent SSC security tools on the RA, ensures its validity and applicability, thus affirming the proposed RA as an effective framework for analysing existing SCM2 solutions and guiding the engineering of new SCM2 systems.


\end{abstract}



\maketitle

\section{Introduction}

In recent years, a meteoric rise in software supply chain (SSC) attacks has occurred. For instance, the 8th Annual State of the Software Supply Chain Report by Sonatype in 2023 \cite{Sonatype2023} reported a 742\% annual increase in SSC attacks in the past three years. An SSC attack combines an upstream attack, where malicious codes are injected into a software artefact via a compromised life cycle activity, and a downstream attack on the consumers who use the artefact \cite{Okafor2022}. Lack of oversight in an artefact's lifecycle makes upstream attacks possible, while a lack of visibility over the distribution and consumption of artefacts opens the door for downstream attacks. Therefore, securing an inventory of software assets requires a thorough and trustworthy visibility over its entire SSC. Such visibility can be provided by \textit{SSC metadata} - the information about an artefact's life cycle, such as how it was constructed and the ``ingredients'' utilised in its construction \cite{Xia2023}. 

The utility of SSC metadata in identifying and mitigating SSC attacks depends on the ability of software security tools to authenticate and interpret this information. Therefore, \textit{a new class of formal SSC metadata that is machine-readable and authenticated has emerged}\cite{Xia2023}. This paper focuses on this SSC metadata class which includes various metadata types such as \textit{Software Bill of Material (SBOM)}, a formal and machine-readable list of ingredients that make up software components \cite{NTIA2021}, \textit{provenance} statements that describe how and by whom a software artefact was produced \cite{SLSA2023} and \textit{software attestation}, which are digitally signed statements about a software artefact \cite{intoto2017}. They are constructed based on a predefined standard (e.g., Software Package Data Exchange (SPDX),\footnote{\url{https://spdx.dev/}} CycloneDX \footnote{\url{https://cyclonedx.org/}}, and in-toto attestation framework\footnote{\url{https://in-toto.io/}}), digitally signed and, optionally, notarised (e.g., 
Supply Chain Integrity, Transparency and Trust (SCITT) \cite{Birkholz2023}) to ensure their authenticity and verifiability. 

Adopting SSC metadata requires practitioners to use software tools to generate, sign, distribute, and consume this information. We denote this software suite as a \textbf{S}oftware Supply \textbf{C}hain \textbf{M}etadata \textbf{M}anagement system \textbf{(SCM2)}. Practitioners face two main challenges when assembling or developing an SCM2: (1) the diversity and overlapping frameworks and terminology related to SSC metadata and (2) the diversity of software solutions making up the design space of an SCM2. As we described above, SSC metadata standards and frameworks vary not only in syntax but also in semantics and content, which are tied to how they define an SSC and the aspects of an SSC they aim to describe (e.g., whether an SSC is a sequence of steps performed by different entities to construct a software artefact \cite{SLSA2023} or a network of providers and consumers of software artefacts \cite{NTIA2021}. These diverse standards and frameworks are actualised by their own toolchains. Moreover, various open-source tools (e.g., Syft\footnote{\url{https://github.com/anchore/syft}}) and proprietary platforms have emerged to address specific aspects of SSC metadata's life cycle, such as generating and signing SSC metadata documents. These tools differ regarding their supported SSC metadata types and standards. They also vary in capability and correctness. The accuracy and completeness of the generated SBOMs caused by tooling competence have been a shared concern among practitioners, according to a recent empirical study about SBOM adoption \cite{Xia2023}. 

The described challenges highlighted the need for a Systematisation of Knowledge (SoK) of SSC metadata to aid practitioners in assembling and developing SCM2 systems. Such a SoK could offer a much-needed comprehensive domain model regarding SSC metadata to facilitate clear communication between architects and stakeholders to identify and define the organisation's information needs regarding SSC and select the suitable types of SSC metadata. An SoK could also provide an architectural blueprint of SCM2 systems, specifying the system's context, scope, and functional decomposition to aid architects in identifying, assessing, and selecting platforms and tools to construct an SCM2. Even though SSC metadata appears in most existing SSC security frameworks (e.g., SLSA \cite{SLSA2023}, SCVS \cite{OWASP2020}, SSF \cite{CNCF2022}), a comprehensive domain model and architectural blueprint of SSC metadata and SCM2 have not been established. 

To this end, this paper presents such an SoK in the form of an \textbf{\textit{Empirically Grounded Reference Architecture (RA)}} for SCM2. An RA can be considered a template for software systems, providing a common vocabulary, taxonomy, architectural vision, and building blocks to guide the development of future systems \cite{Muller2008}. RAs have traditionally been constructed to provide SoK and guidance for Service-oriented computing \cite{Arsanjani2007}, Cloud computing \cite{Liu2011}, Big data \cite{Paeaekkoenen2015}, and Internet of Things \cite{Weyrich2016}, where a diverse and decentralised set of stakeholders are required to interoperate and cooperate. Today's SSC metadata management exhibits similar multiplicity due to the diversity of frameworks, standards, tool sets, and stakeholders, making an empirically grounded RA a suitable solution.  

Systematically designing  RAs through an empirically grounded approach facilitates capturing the key concepts and elements of the state-of-the-art frameworks and architectures, thus improving the validity and reusability of the RA \cite{heitmann2011, Galster2011}. Hence, we follow the empirically grounded RA design methodology proposed by Galster and Avgeriou \cite{Galster2011}, which includes two aspects for the systematic design of RAs: use of an empirical foundation where practice-proven concepts and building blocks are used for constructing an RA and validity of the constructed RA where the proposed RA is evaluated in terms of its correctness and applicability. As SSC metadata management is a novel and compelling area of interest among software practitioners \cite{Xia2023}, we use state-of-the-art industry-driven and peer-reviewed SSC security frameworks and architectures to establish a solid empirical foundation for the RA. Our proposed RA for SCM2 comprises two components: (1) the \textit{domain model} (Section \ref{sec:domain_model}) presents the key concepts related to SSC metadata and the relationships between them, accompanied by the life cycle of SSC metadata and the related actors. The \textit{architectural blueprint} describes the scope and positioning of SCM2 within an SSC (Section \ref{sec:context_model}) and the function units making up an SCM2 instance (Section \ref{sec:container_model}). To ensure the validity of the proposed RA, we evaluate its correctness and utility by mapping the SCM2 elements of five prominent SSC security tools onto the concepts and architecture of the RA. Based on the architectural mapping, we discussed the state of practice of SCM2 and presented a reference instantiation of SCM2 from the existing open-source tools.

\section{Methodology}

Our methodology was developed based on the systematic design approach proposed by Galster and Avgeriou \cite{Galster2011} and comprised six steps (see Figure \ref{fig:methodology_overview}):

\vspace{-2.5mm}
\begin{figure}[ht]
	\centering
	\includegraphics[width=0.5\textwidth,,trim={0cm 2.5cm 0cm 0cm},clip]{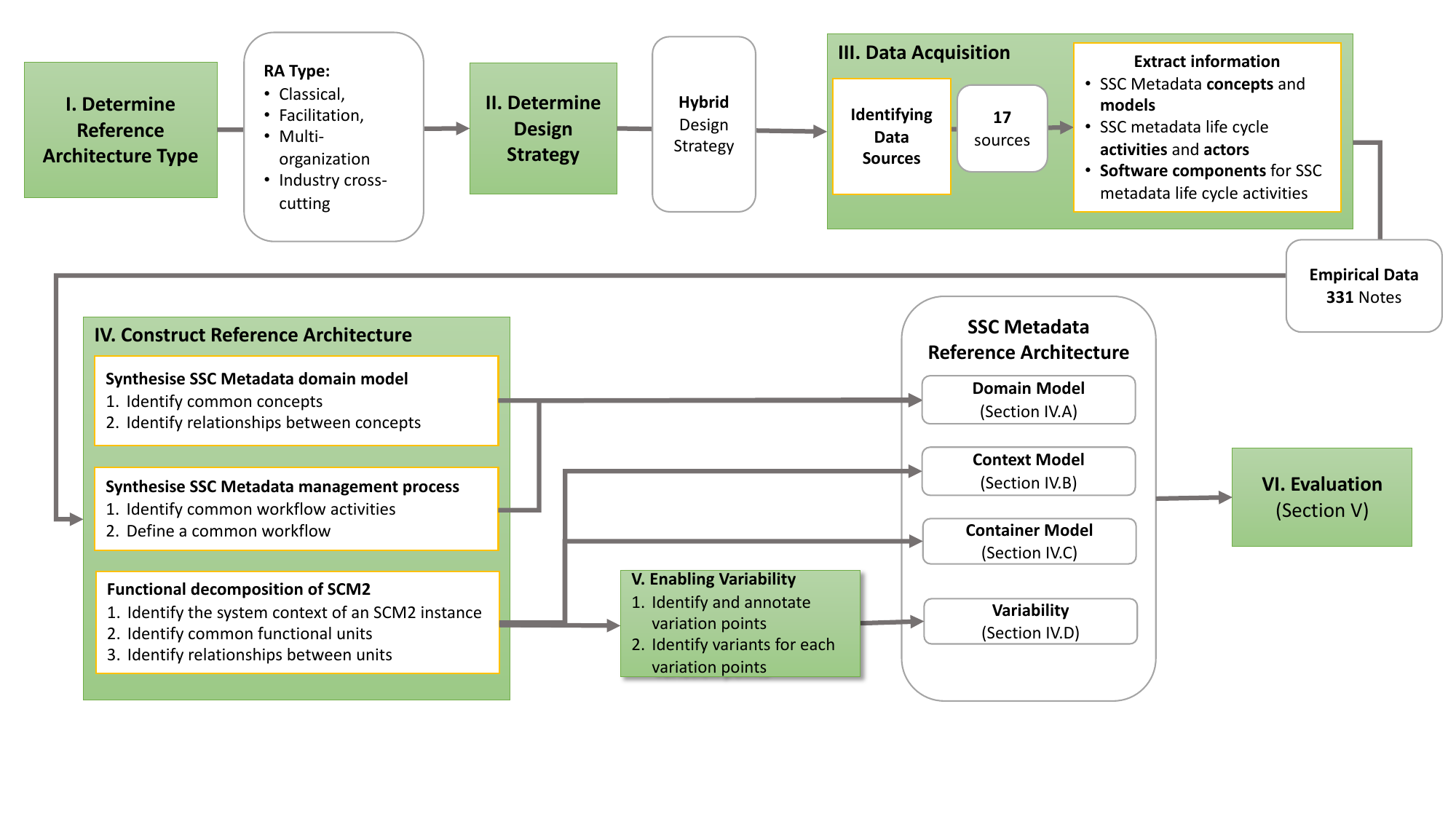}
     \vspace{-3.5mm}
	\caption{Methodology for constructing the reference architecture for SSC Metadata life cycle management}
	\label{fig:methodology_overview}
\end{figure}

 \vspace{-2.5mm}
\noindent \textbf{Step 1 - Determine RA type:} This crucial decision determines the design strategy and data collection method for constructing an RA. Based on Angelov's classification system \cite{Angelov2009}, we classified our RA as a \textit{classical}, \textit{facilitation} RA intended for use by \textit{multiple organizations} in an \textit{industry cross-cutting context}. This means that our RA would be based on experiences from preexisting systems, with the primary goal of facilitating the design and development of new systems. It would be used by multiple organizations, both within and outside the software industry, that require access to SSC metadata.

\vspace{1.5mm}
\noindent \textbf{Step 2 - Determine Design Strategy:} 
A \textit{hybrid design strategy} was adopted, meaning both industry-driven SSC security frameworks and peer-reviewed academic articles were utilised as inputs for the construction of the proposed SCM2 RA.

\vspace{1.5mm}
\noindent \textbf{Step 3 - Data Acquisition:}
The first activity in this step was \textit{identifying the data sources for constructing the empirically grounded RA}. A preliminary set of SSC security frameworks and architectures were identified based on the authors' experiences with and discussions within SSC security industry projects. This preliminary selection was expanded using the snowballing sampling technique \cite{wohlin2014}. If a framework had multiple revisions, only the latest version was selected. This selection process resulted in 13 industry-driven reports and standards and four peer-reviewed articles which, together, specify \textbf{eleven SSC security frameworks} that served as the empirically grounded data input for our RA. Table \ref{tab:relatedworks} presents an overview of the chosen frameworks, focusing on their coverage of SSC metadata and its life cycle. We classify the coverage into four classes: ``0'' denotes that a framework does not offer any details about an SSC metadata life cycle stage, ``1'' indicates that it provides an overall description, ``2'' indicates that it describes workflow activities and actors, and ``3'' denotes that it provides functional decomposition and describes involved software components. For brevity, we make the details of these frameworks available in the online appendix of the paper (see Section \ref{sec:data}).

\renewcommand{\arraystretch}{1.0}
\begin{table*}[ht] \footnotesize
\centering
\caption{SCM2 Support of existing SSC Security Frameworks}
  \label{tab:relatedworks}
\vspace{-2.0mm}
\resizebox{0.8\linewidth}{!}
{\begin{tabular}{p{0.25\textwidth} c c p{0.15\textwidth} c c c}
\toprule
\textbf{Framework/} & \textbf{Scope} & \textbf{Supported SSC} & \textbf{Supported SSC}  & \multicolumn{3}{c}{\textbf{Metadata Life Cycle Coverage}} \\
\cline{5-7}
\textbf{Architecture} & & \textbf{Metadata Types} & \textbf{Metadata Standards} & \textbf{Generation} & \textbf{Sharing} & \textbf{Consumption} \\
\midrule
SEI SCRM \cite{SEI2010,SEI2010a,Alberts2011} & SSC Concepts & N/A & N/A & N/A & N/A & N/A \\

In-toto Attestation Framework & SSC Metadata & Attestation & In-toto Attestation  & 3 & 1 & 3 \\
 \cite{intoto2017} & & Provenance (Partially) & Specification &  &  &  \\

Software Delivery Governance \cite{Singi2019} & SSC Metadata & Provenance & N/A & 1 & 3 & 1 \\

SCVS \cite{OWASP2020} & SSC Metadata & SBOM & SPDX & 1 & 0 & 1 \\

SBOM Generation Playbook \cite{NTIA2021,NTIA2021h,NTIA2021e,NTIA2021p1}& SSC Metadata & SBOM & Any & 2 &  1& 1  \\

SBOM Sharing Playbook \cite{NTIA2021g,Stoddard2023} & SSC Metadata & SBOM & Any &  0 & 2 & 1 \\

SBOM Consumption Playbook \cite{NTIA2021p2} & SSC Metadata & SBOM & Any &  0 & 1 & 2 \\

SBOM Tool Classification \cite{NTIA2021j} & SSC Metadata & SBOM & Any & 1 & 0 & 1 \\

SSF Architecture \cite{CNCF2022} & SSC Metadata & Attestation & Any & 3 & 3 & 0 \\

SCITT \cite{Birkholz2023} & SSC Metadata & Attestation & SCITT Transparent Statement & 1 & 3 & 2 \\

SLSA \cite{SLSA2023} & SSC Metadata & Provenance & SLSA Provenance Model,  & 1 & 1 & 1 \\
 & & Attestation & SLSA Software Attestation Model &  &  &  \\
\bottomrule

\end{tabular}}
\end{table*}

The second activity was \textit{extracting information} from data sources using the document analysis method \cite{bowen2009}. The authors independently analysed the sources and extracted relevant information regarding SSC metadata, SSC metadata life cycle activities and actors, and software components introduced to carry out SSC metadata life cycle activities. The extracted information were captured in \textbf{331 digital note cards} and stored in a version-controlled repository. The authors incrementally review the note cards and organised them into a knowledge graph in weekly discussion sessions. We leveraged Obsidian\footnote{\url{https://obsidian.md}}, a personal knowledge base system, as the tool to capture digital notes and construct the knowledge graph. The resulting knowledge graph can be found in the online appendix of the paper (see Section \ref{sec:data}).

\vspace{1.5mm}
\noindent \textbf{Step 4 - Reference Architecture Construction:}
The RA consists of a domain model and an architectural blueprint. The domain model comprised (1) \textit{a static model} showing concepts related to SSC metadata and how they relate with each other and (2) \textit{a dynamic model} that describes the life cycle of SSC metadata. These models were derived from the common concepts and activities captured in the knowledge graph. 

The architectural blueprint for SCM2 was constructed from the \textit{union} of architectural components captured in the knowledge graph. Union rather than intersection was chosen because this approach extends the coverage of the constructed RA. This design was driven by our observation that none of the existing frameworks provides a complete coverage across SSC metadata types and life cycle activities. The constructed RA was documented using two higher-level abstractions defined by the \textit{C4 architectural view model} \cite{Brown2018}. Particularly, we utilised the system context model to describe how an SCM2 instance fits in to the overall IT environment and the container model to describe the functional units within an SCM2 instance.

\vspace{1.5mm}
\noindent \textbf{Step 5 - Enabling Variability:} 
Instantiating software architecture from an RA necessarily introduces variations such as the modification or removal of functional units. An RA can support variability by specifying the variation points (architectural elements where changes might occur) and variants (the possible options for those variation points) \cite{Galster2011}. We identified the variation points by searching for SSC metadata life cycle activities and functional units where the analysed frameworks and architectures diverged. The architectural components and activities utilised by these frameworks and architectures represent the variants. We used textual annotation to describe the variability of the proposed RA.

\vspace{1.5mm}
\noindent \textbf{Step 6 - Evaluation:}
To evaluate the correctness and usefulness of the empirically grounded RA, we follow the architectural mapping-oriented RA evaluation approach proposed in \cite{Galster2011} and \cite{Liu2023}. The evaluation involves mapping prominent SSC security solutions to the proposed RA and analysing how they comply. We selected 5 security tools available for the use by practitioners and widely recognized for management of SSC metadata. We will discuss detailed evaluation methodology in Section \ref{sec:evaluation}.

\section{SCM2 Reference Architecture}

\subsection{Domain Model}
\label{sec:domain_model}

\subsubsection{SSC metadata Concepts}

\begin{figure}[ht]
    \includegraphics[width=\linewidth]{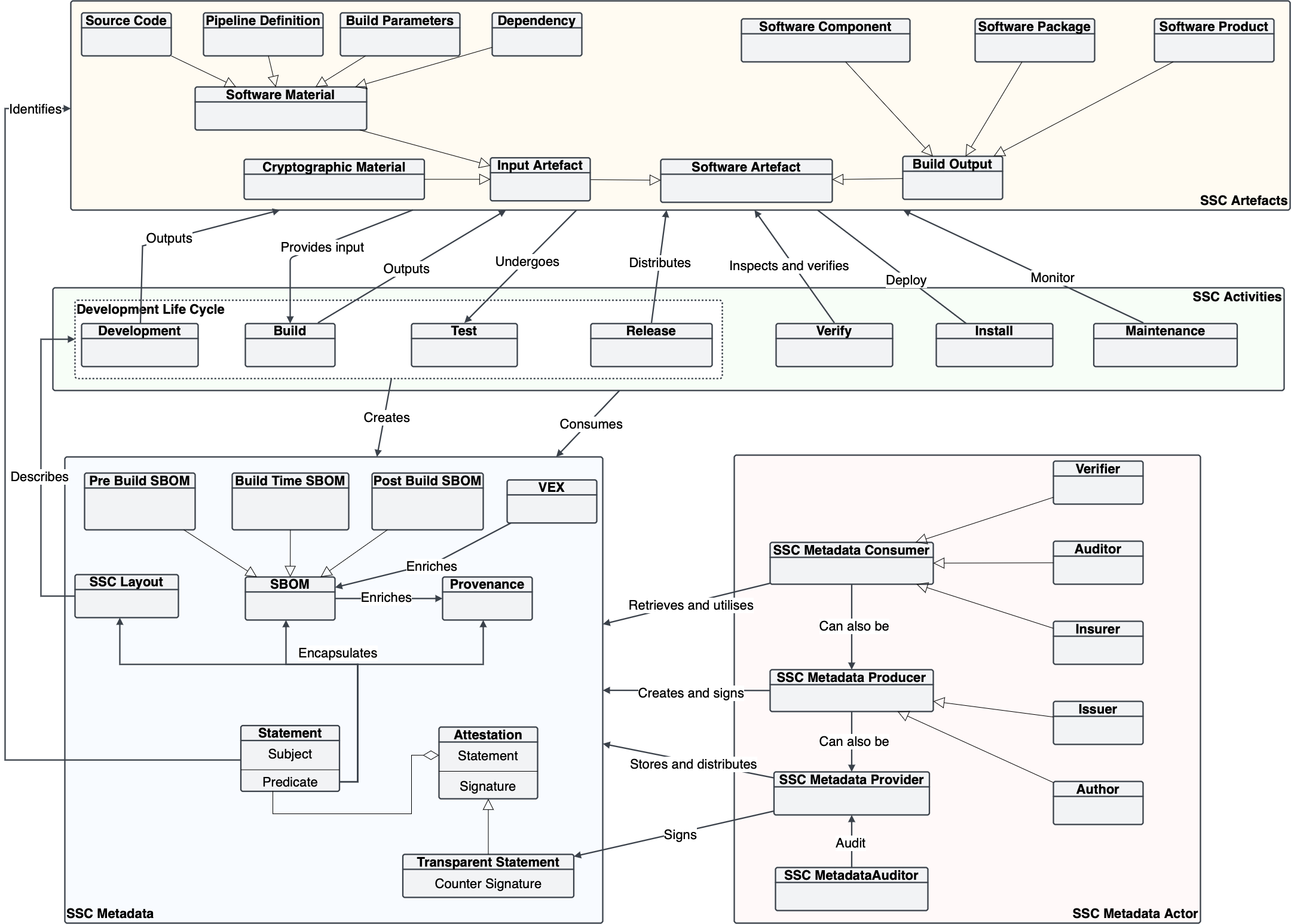}
    \caption{Domain Model for SSC and SSC Metadata Concepts}
    \vspace{-4.5mm}
    \label{fig:SSC_domain_model}
\end{figure}

We organised the concepts relevant to SSC metadata and the relationship between them into four clusters (Figure \ref{fig:SSC_domain_model}), representing four perspectives to define SSC metadata: the \textit{SSC activities} that create and consume metadata, the \textit{SSC artefact} that metadata describes, the \textit{SSC metadata actors} who act upon the metadata, and finally the taxonomic relationships among SSC metadata types themselves. 

\vspace{1.5mm}
\noindent \textit{SSC Activities:} An SSC can be considered a series of coordinated \textit{SSC activities} performed by a network of actors to create and distribute software products to end-users \cite{Singi2019, Okafor2022, intoto2017, SLSA2023, CNCF2022}. On the producer side, SSC activities include \textit{development}, \textit{build}, \textit{test}, and \textit{release}. The activities on the consumer side after retrieving a software product include \textit{verification} (e.g., checking digests and verifying the product against its metadata), \textit{installation}, and \textit{maintaining} the software product throughout its life span. 

\noindent \textit{SSC artefacts:} SSC activities consumes and produces \textit{software artefacts}. The existing SSC security frameworks defined artefacts in various ways, from any ``immutable blob of data'' \cite{SLSA2023} to ``item that is moving along the supply chain'' \cite{Birkholz2023} to the principal output of a secure software factory that downstream users would consume \cite{CNCF2022}. Our domain model classifies SSC artefacts into input and output artefacts. SSC activities consume input artefacts include \textit{software materials} \cite{intoto2017} such as \textit{source code}, build \textit{pipeline definition}, external \textit{build parameters} \cite{SLSA2023} and \textit{dependency}, which are other artefacts that need to be collected before building a software \cite{CNCF2022}. The input artefacts also include cryptographic material such as certificates, tokens, and signing keys \cite{CNCF2022}. The output artefact, also known as \textit{build output}, contains the output produced by a build process \cite{SLSA2023}. \textit{Software components}, the focus of SBOM-focused reports by NTIA \cite{NTIA2021} and Core Infrastructure Initiative \cite{Nagle2020}, can be considered fine-grained build outputs that another software can call. \textit{Software packages}, the focus of SLSA framework \cite{SLSA2023}, can be a collection of software components published for others. A \textit{software product} \cite{intoto2017} can be considered a synonym or a super-set of software packages. 

\vspace{1.5mm}
\noindent \textit{SSC metadata} provides machine-readable and authenticated descriptions of SSC artefacts and activities. At the core of an SSC metadata document is a collection of \textit{statements}, consisting of a set of \textit{predicates} about a \textit{subject} such as an SSC activity or artefact \cite{intoto2017}. The predicates can describe various types of information:
\begin{itemize}[left=0pt]
    \item \textit{Software Bill of Material (SBOM)} is a machine-readable nested inventory of components within a package or product \cite{NTIA2021}. SBOM can be written in interoperable formats such as SPDX, CycloneDX, and SWID. SBOM can be further enriched by \textit{Vulnerability Exploitability eXchange (VEX)} documents, which integrate vulnerability information with component data from SBOM \cite{Stoddard2023,CNCF2022,NTIA2021}. 
    \item \textit{Provenance} describes how an artefact was produced. It can be considered a claim that some entity produced one or more artefacts by executing a build pipeline definition according to some parameters, possibly using other artefacts as input \cite{SLSA2023}. An alternative definition by the SCVS framework \cite{OWASP2020} considers provenance as the origin and chain of custody of a software component. 
    \item \textit{SSC Layout} is a document providing an ordered list of steps, requirements for the steps, and actors responsible for each step that is required to be carried out in the SSC to create a software product \cite{intoto2017}.
\end{itemize}

\textit{Software attestation} is generated when a digital signature authenticates the statements. By signing an attestation, the authors of an SSC metadata document claim that the predicates (e.g., SBOM, provenance) about a subject (e.g., a software package) are factual \cite{SLSA2021}. A trusted authority can countersign an attestation to prove its existence and validity. A countersigned attestation can be denoted as a \textit{transparent statement} \cite{Birkholz2023}. 

\vspace{1.5mm}
\noindent \textit{SSC Metadata Actors} carry out life cycle activities of SSC metadata documents. They can be organised into three classes.
\begin{itemize}[left=0pt]
    \item \textit{SSC Metadata Producers} create and sign SSC metadata. Metadata producers can be project owners \cite{intoto2017} or package producers \cite{SLSA2023}, who are responsible for defining the layout of the supply chain of a software artefact. Metadata producers can also be functionaries (e.g., developers, build systems, etc.) that handle individual steps in the life cycle of a software artefact \cite{intoto2017}. SSC metadata producers are also known as \textit{issuer} \cite{Birkholz2023} and \textit{author} \cite{NTIA2021p1,Stoddard2023}.
    \item \textit{SSC Metadata Consumers} retrieve and utilise SSC metadata \cite{SLSA2023, Stoddard2023, NTIA2021p2}.  A \textit{verifier} is a consumer who leverages SSC metadata to verify the authenticity and integrity of software artefacts \cite{SLSA2023,intoto2017,Birkholz2023}. The inspection process of the artefacts also creates more metadata, such as VEX. Other types of consumers include \textit{insurers} and \textit{auditors} \cite{NTIA2021a} who leverage SSC metadata to assess an organisation for regulatory or insurance purposes. 
    \item \textit{SSC Metadata Providers} \cite{Birkholz2023} operate the intermediary services that bridge SSC metadata producers and consumers. Providers can also act as a notary, vouching for the existence and the correctness of SSC metadata documents registered by producers \cite{Birkholz2023}. A \textit{SSC metadata auditor} is a particular type of auditor who assesses the trustworthiness of the providers based on the log or provenance of the SSC metadata documents distributed by providers \cite{Birkholz2023}.
\end{itemize}

\subsubsection{SSC Metadata Life Cycle} \label{lifecycle}

\begin{figure*}[!ht]
    \includegraphics[width=\linewidth,,trim={0cm 0.5cm 0cm 0cm},clip]{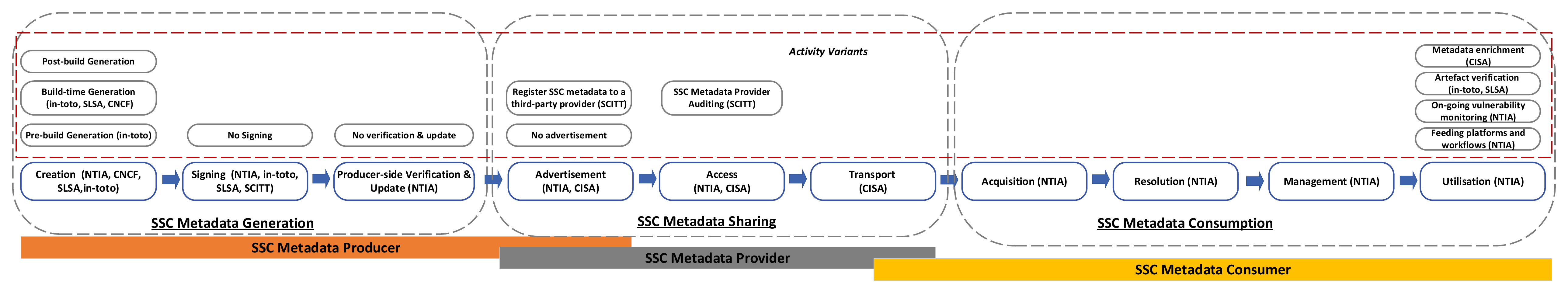}
    \caption{Life cycle of SSC Metadata}
    \label{fig:SSC_metadata_life_cycle}
    \vspace*{-2.0mm}
\end{figure*}

The life cycle of SSC metadata comprises ten activities that can be organised into generation, sharing, and consumption phases. Figure \ref{fig:SSC_metadata_life_cycle} depicts the life cycle activities and the actors handling those activities. Where the analysed frameworks and architectures diverge, we present their alternative activities as variations. 


\vspace{1.5mm}
\noindent \textit{SSC Metadata Generation Phase} consists of 3 main activities starting with \textit{SSC metadata creation} by individuals or automated tools. The supplier playbook by NTIA \cite{NTIA2021p1} defines a three-step process that includes identifying software components for delivery, acquiring their data, and capturing the data in an SBOM format. NTIA \cite{NTIA2021} further prescribes that SBOM authors must include SBOMs of all the dependencies or create those SBOMs on a best-effort basis should such information be unavailable. The in-toto attestation framework \cite{intoto2017} prescribes that all ``functionaries'', actors participating in the creation of software artefacts, must record metadata after finishing their activity. The SLSA 1.0RC specification \cite{SLSA2023} also requires that build processes automatically generate provenance attestation describing how the artefact was built (by whom, using which process or command, with which input artefacts). Similarly, the Secure Software Factory framework \cite{CNCF2022} also prescribes automated metadata generation carried out by a software component embedded into build pipelines.

There are three variants of carrying out the SSC metadata creation: \textit{pre-build} generation, \textit{build-time} generation, and \textit{post-build} generation \cite{NTIA2021p1}. Pre-build generation activity creates metadata about source code, such as static dependency declaration in the form of \texttt{requirements.txt} in Python projects and \texttt{package.json} in NodeJS projects. The SSC layout introduced by the in-toto framework \cite{intoto2017} is also a product of a pre-build generation activity. The build-time generation activity generates SSC metadata during or before the end of the build process. It is the most common variant of SSC metadata creation activity, featured in most frameworks, including in-toto \cite{intoto2017}, SLSA 1.0RC \cite{SLSA2023}, and Secure Software Factory \cite{CNCF2022}. The final variant is the post-build generation, which captures metadata such as VEX from automated tests and auditing activities performed on the newly built artefacts \cite{NTIA2021}.

The second activity in the generation phase is \textit{signing} the generated SSC metadata \cite{NTIA2021p1}. This activity is required by the in-toto framework \cite{intoto2017} and the SCITT architecture \cite{Birkholz2023}. The second and third levels of the build track of SLSA 1.0RC specification \cite{SLSA2023} also require SSC metadata to be digitally signed. It should be noted that signing SSC metadata is not mandatory. For instance, the supplier handbook \cite{NTIA2021p1} and level one of the build track of SLSA 1.0RC consider metadata signing an optional activity. Therefore, we introduce \textit{``no signing''} as a variant of the signing activity.

The third activity is \textit{verifying and updating} the content of the created metadata. This activity was introduced in the NTIA's supplier playbook \cite{NTIA2021p1} as a remedy for potential errors due to the SBOM creation software tools and errors inherited from upstream SBOMs. Interestingly, we could not find the provider-side verification and metadata update in other frameworks and architectures. Therefore, we introduce \textit{``no verification and update''} as a variant. 

\vspace{1.5mm}
\noindent \textit{SSC Metadata Sharing Phase} bridges the SSC metadata producers and consumers \cite{NTIA2021g,Stoddard2023}. The first activity in this phase is \textit{``advertisement''}, which SSC metadata producers perform to inform consumers of the availability of SSC metadata documents and how to access them \cite{NTIA2021g}. Producers also use the advertisement to notify consumers of corrections or updates of a previously published SSC metadata document. Producers can handle the advertisement themselves or \textit{register SSC metadata documents with a third party provider} for distribution \cite{NTIA2021g, Birkholz2023}. Alternatively, producers can \textit{omit the advertisement}. In this case, consumers would query producers for SSC metadata of the received software artefacts \cite{Stoddard2023}. 

The second activity is \textit{access}, in which consumers' requests for SSC metadata are authorised according to some policies specified by the producer. These policies are defined in a fine-grained manner, limiting access to specific versions or potions of SSC metadata documents \cite{Stoddard2023}. The responsibility for assessing and enforcing access policies lies with either producers or providers who host the SSC metadata documents. If a third-party provider is utilised, the \textit{provider auditing} activity could be performed \cite{Birkholz2023}.  

The metadata sharing phase concludes with \textit{transport} SSC metadata documents from a producer or a provider to the authorised consumers using a secure mechanism \cite{NTIA2021g, Stoddard2023}. According to \cite{Stoddard2023}, transfer can be done from single point to single point or single point to multi point based on the targeted consumers.

\vspace{1.5mm}
\noindent \textit{SSC Metadata Consumption Phase} comprises four activities \cite{NTIA2021p2}. The first activity is the \textit{acquisition} of SSC metadata documents via the sharing mechanisms provided by producers or providers. The second activity is \textit{resolution}, in which a consumer maps subjects of the statements in an SSC metadata document onto software artefacts within their inventory to establish a link between them. This step might also involve resolving transitive dependency identified in an SBOM. 

The third activity is applying \textit{content management} technique on the acquired SSC metadata documents, such as defining and enforcing policies about their storage, data retention and life cycle. The consumption phase ends here if the SSC metadata documents are not utilised in any enterprise and IT workflow. Otherwise, the last activity is \textit{utilisation} where the acquired SSC metadata documents are \textit{fed into platforms or workflows} (e.g., Application Security Posture Management, Software Asset Management), utilised for \textit{on-going vulnerability monitoring}, employed for in artefact verification processes \cite{intoto2017,SLSA2023}. An organisation can also \textit{enrich} an acquired SSC metadata document, such as by appending their own VEX, and share it with other consumers \cite{Stoddard2023}. 

\subsection{Architectural Blueprint}

This section presents the architectural blueprint for the SSC Metadata Management System (SCM2), which is a suite of software tools for SSC metadata actors (defined in section \ref{sec:domain_model}) to carry out the life cycle activities (defined in section \ref{lifecycle}). We construct the architectural blueprint based on the extracted empirical data and present the architectural blueprint using two models: Context Model and Container Model.

\subsubsection{Context Model}
\label{sec:context_model}

\begin{figure}[!ht]
    \centering
    \includegraphics[width=1\linewidth,,trim={0cm 0cm 0cm 0cm},clip]{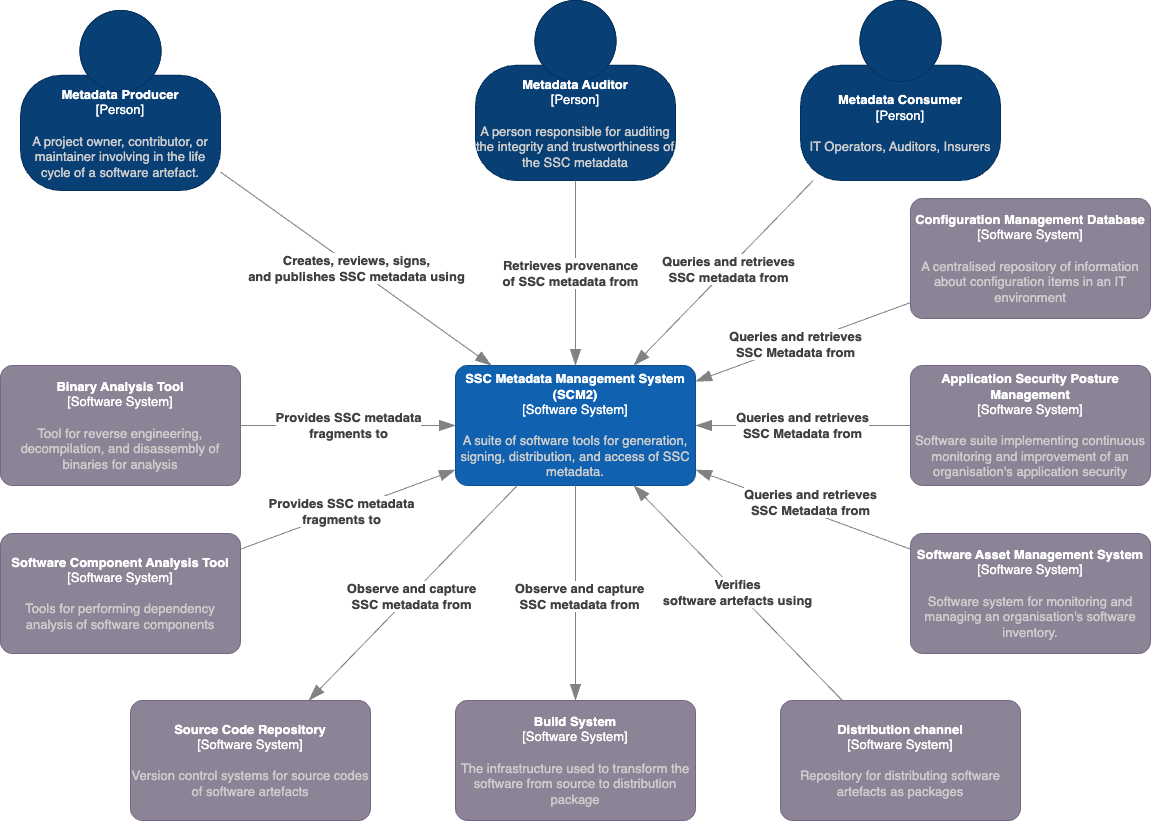}
    \caption{Context model of SCM2}
    \label{fig:SCM2_Context}
    \vspace*{-2.0mm}
\end{figure}

The system context diagram in Figure \ref{fig:SCM2_Context} presents the positioning and scope of SCM2, reflected via its interactions with three types of actors and eight types of external systems.


The actors and external systems interacting with SCM2 can be organised into two groups, namely producer and consumer. The producer group contains the metadata producers, binary analysis tools, software component analysis tools (SCA), source code repositories, and build systems. From the perspective of SCM2, the first three actors and external systems are push-based producers because they push fragments of SSC metadata to SCM2 for constructing SSC metadata documents. On the other hand, source code repositories and build systems are pull-based producers because SCM2 observes and pulls metadata fragments from these systems when constructing SSC metadata document for an artefact. 

The SSC metadata generation process of SCM2 can be initiated in two ways. Firstly, it can be triggered by the build process as a part of a CI/CD pipeline. This workflow is suggested by many frameworks \cite{SLSA2023, OWASP2020, CNCF2022} and particularly relevant to the scenarios where the SSC metadata producer of a software artefact is also the developer and project owner of an artefact. The second approach is to have the process triggered manually by SSC metadata producer, when the need for SSC metadata arises, such as to document their software inventory with SBOM documents. This workflow is particularly relevant to organisations that primarily consume rather than produce software artefacts. The producers might rely on SCA and binary analysis tools to extract the necessary information about their software artefacts. Due to the separation of concern, we decided to place SCA and binary analysis tools outside the scope of SCM2. It means that SCM2 does not contain the ability to perform composition analysis or reverse engineering binaries. It only contains the ability to generate SSC metadata document from the output of these tools. 

The second group of actors and external systems of SCM2 consists of metadata consumers, configuration management databases, application security posture management systems, software asset management systems, distribution channels, and metadata auditors. All actors and external systems in this group besides auditors query and retrieve SSC metadata documents from SCM2 for various purposes, such as verifying an incoming software artefact before installation. The metadata auditor, on the other hand, consumes the provenance of the SSC metadata documents in order to assess the trustworthiness of the SSC metadata and the SCM2 itself. 

\subsubsection{Container Model}
\label{sec:container_model}

\begin{figure*}[!ht]
    \centering
    \includegraphics[width=\linewidth,,trim={0cm 0cm 0cm 0cm},clip]{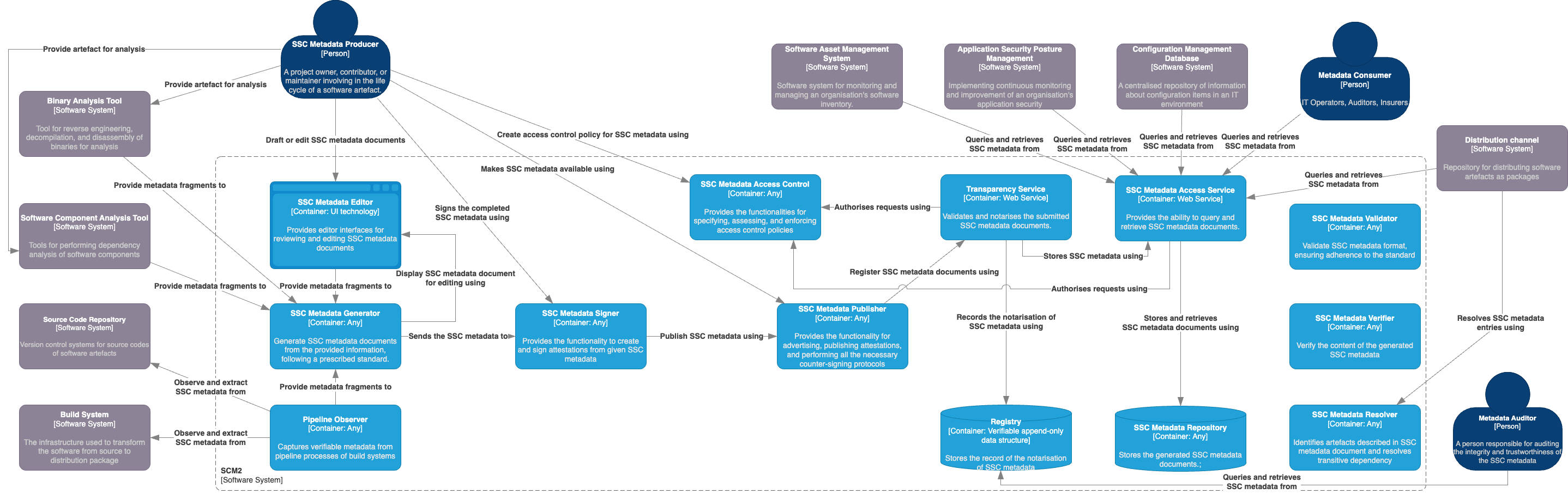}
    \caption{Container model of SCM2}
    \label{fig:SCM2_Container}
    \vspace*{-2.0mm}
\end{figure*}

The container diagram in Figure \ref{fig:SCM2_Container} presents the distribution of the functionality of an SCM2 system across 13 containers. It should be noted that a container in the C4 architectural view model denotes a separately runnable unit that executes code or stores data rather than a software container (e.g., Docker). We defer the binding of containers to specific technologies to the design time when architects instantiate concrete SCM2 architecture from the RA. The containers of an SCM2 system organised into five groups according to the SSC metadata life cycle phase in which they manage.

\vspace{1.5mm}
\noindent\textbf{SSC metadata generation} group contains three containers. \textit{SSC Metadata Editor} provides a graphical interface for reviewing and editing SSC metadata document. This tool belongs to the type ``Edit'' of the category ``Produce'' in the tool classification taxonomy by NTIA \cite{NTIA2021j}. It supports SSC metadata producers in authoring pre-build metadata (e.g., in-toto's SSC layout metadata \cite{intoto2017}), supplying post-build metadata and verifying the auto-generated SSC metadata documents \cite{NTIA2021p1}. 

\textit{Pipeline Observer} captures information about tasks in a build pipeline, such as fetching source code and dependencies, building and testing artefacts. The captured evidence is the input for generating SSC metadata documents. Pipeline observer was introduced in the Secure Software Factory reference architecture \cite{CNCF2022}.

\textit{SSC Metadata Generator} creates SSC metadata document from the input information according to a prescribed standard such as SPDX \cite{NTIA2021j}. It should be noted that most existing tools combine the metadata generator with a Software Composition Analysis tool or a pipeline observer (e.g., by integrating the generator tool into a build pipeline). In this reference architecture, we model the metadata generator as a separate function unit to follow the separation of concern principle and facilitate the reuse of the software logic for generating SSC metadata document for different standards and templates. 

\vspace{1.5mm}
\noindent\textbf{SSC metadata signing} group consists of one container: \textit{SSC Metadata Signer}. The signer provides metadata producers with the ability to sign the generated SSC metadata documents, effectively turning them into authenticated statements about a software artefact (software attestations). For example, the ``cosign'' component\footnote{\url{https://github.com/sigstore/cosign}} of the ``sigstore'' project can be used as an SSC metadata signer by utilising its built-in support for the in-toto attestation framework \cite{intoto2017}.

\vspace{1.5mm}
\noindent\textbf{SSC metadata publishing} group comprises two containers. \textit{SSC Metadata Publisher} provides metadata producers with the functionality for publishing and advertising the generated SSC metadata documents. This functionality reflects and implements the ``advertisement'' phase described in the SBOM sharing life cycle \cite{NTIA2021g,Stoddard2023}. The publisher can also implement the notarising workflow described by the SCITT architecture \cite{Birkholz2023} where an external authority verifies and countersigns a submitted SSC metadata document to attest for its trustworthiness. The client-side of sigstore's rekor\footnote{\url{https://github.com/sigstore/rekor}} can be utilised to implement the publisher logic. 

\textit{SSC Metadata Access Control} provides metadata providers with the functionality for specifying access control policies of the SSC metadata documents that they publish. It also assesses and authorises requests for SSC metadata documents. This container was introduced based on the requirement for access control mechanism outlined in the SBOM sharing life cycle \cite{NTIA2021g,Stoddard2023}.

\vspace{1.5mm}
\noindent\textbf{SSC metadata sharing group} consists of four containers. \textit{SSC Metadata Repository} stores the generated SSC metadata documents. This container corresponds to the ``artefact repository'' component in the Secure Software Factory reference architecture \cite{CNCF2022}. \textit{SSC Metadata Access Service} acts as a wrapper around the metadata repository, providing the metadata consumers and external systems with the ability to query and retrieve SSC metadata document. This container implements the ``access'' and ``transport'' phases of the SBOM sharing life cycle \cite{NTIA2021g,Stoddard2023}. It relies on the SSC metadata access control container to assess and authorise incoming requests. 

\textit{Registry} is a verifiable, append-only data store for the records regarding the notarisation of SSC metadata documents. This container was introduced in the SCITT architecture \cite{Birkholz2023}. The public instance of Rekor\footnote{\url{https://rekor.sigstore.dev}} represents an example of a registry. \textit{Transparency Service} validates and notarises the submitted SSC metadata documents and record the process in the registry. This container was introduced in the SCITT architecture \cite{Birkholz2023}.

\vspace{1.5mm}
\noindent\textbf{SSC metadata consumption} group consists of three containers. \textit{SSC Metadata Validator} provides the ability to validate the syntax of SSC metadata documents to ensure their adherence to a given standard. This component was introduced by the \cite{NTIA2021p2}. \textit{SSC Metadata Verifier} provides the ability to verify the information contained within a SSC metadata document. This component was introduced by the \cite{NTIA2021p2}. \textit{SSC Metadata Resolver} implements the resolution step introduced in \cite{NTIA2021p2}. It identifies artefacts described in SSC metadata document and resolves transitive dependency.

\vspace{-3.5mm}

\subsection{Variability}
\label{sec:variability}

This section discusses the variability of SCM2 and presents some prominent variants in terms of the selection, deployment, governance of containers that practitioners can choose when instantiating a concrete SCM2 system for their use case. For brevity, we present variation points and variants in five groups that correspond to architectural design decisions as follows. 

\vspace{1.5mm}
\noindent \textbf{Monolithic vs Distributed:} This design decision impacts the deployment and interaction between containers making up an SCM2 instance. The result of this decision leads to two SCM2 variants which we denote as \textit{monolithic} and \textit{distributed}. A \textit{monolithic SCM2} operates and scales as a single unit. For instance, the system might contain a core framework that orchestrates SSC metadata life cycle activities and exposes hooks for integration with modules that implement the functionalities of the containers in Figure \ref{fig:SCM2_Container}. The framework and its modules might run within a process or a software container. Scaling of such monolithic SCM2 might be achieved by creating replicas of the system and placing them behind a load balancer. 

On the other hand, a \textit{distributed SCM2} comprises multiple units that can operate autonomously and be deployed on separated machines. The containers described in Figure \ref{fig:SCM2_Container} can be mapped directly onto these units. The microservice architecture can be applied to design the architecture of the distributed SCM2 system.

\vspace{1.5mm}
\noindent \textbf{Centralised vs Decentralised:} This design decision reflects the relationship between the entities owning and operating a SCM2 system. It influences the security requirements of interaction between SCM2's containers and the need for decentralisation of the SSC metadata registry and repository. A \textit{centralised SCM2} system operates within the trust domain of the system owner. On the other hand, a \textit{decentralised SCM2} system spans multiple trust domains. For instance, the system might have multiple separated SSC metadata producers and SSC metadata providers. In this context, producer's containers cannot trust provider's containers implicitly, and therefore authentication and authorisation mechanisms must be introduced. The SSC metadata registry and repository might need to be replicated and maintained by the consensus of a group of SSC metadata providers (e.g., by leveraging a distributed ledger) to mitigate the lack of trust in a single entity. 

\vspace{1.5mm}
\noindent \textbf{Formal metadata distribution vs Informal sharing:} This design decision reflects the degree of separation between SSC metadata producers and SSC metadata consumers. When consumers are also producers, such as when an organisation applies a software composition analysis tool to construct SBOM for their software inventory, containers related to the advertisement, access, and transport of SSC metadata can be omitted. We also denote this variant as \textit{single-user SCM2} system. When consumers and providers are separate entities connected via a supply chain relationship, formal distribution and access control mechanisms are necessary. We denote this variant as \textit{multi-user SCM2} system. 

\vspace{1.5mm}
\noindent \textbf{Signing:} This design decision reflects the maturity of SSC security procedures and SSC metadata adoption within an organisation. For instance, the signing of SSC metadata is only required from level two requirement of both the build track of SLSA \cite{SLSA2023} and the SBOM control family of SCVS \cite{OWASP2020}. We denote the variant of SCM2 without the SSC metadata signer container as \textit{no-sign SCM2} system.

\vspace{1.5mm}
\noindent \textbf{Notarisation:} This design decision reflects the maturity of the SSC metadata sharing mechanism. Notarisation includes the countersigning of SSC metadata documents performed by a trusted authority and the recording of such countersigning. For instance, the SCITT architecture \cite{Birkholz2023} presents a notarisation protocol. An \textit{unnotarised SCM2} system omits the transparency service and the registry, sending the published SSC metadata documents to the repository directly. 

\section{Evaluation}
\label{sec:evaluation}

The correctness and utility of the SCM2 RA were evaluated by mapping the SSC security functionality and components of prominent SSC security solutions 
onto the RA's architectural constructs. It should be noted that since the SCM2 RA was an synthesis of the existing standards and frameworks governing SSC security and SSC metadata, these concrete SSC security solutions and services were not the inputs to the RA and thus can be utilised to test the RA itself. The evaluation aimed at two following research questions:

\begin{itemize}[left=0pt]
    \item \textbf{RQ1:} Can the architectural constructs of SCM2 RA describe the functionality and architecture of existing SCM2 solutions?
    \item \textbf{RQ2:} Where do existing solutions focus or deviate from the SCM2 RA?
\end{itemize}

\subsection{Methodology}

We selected SSC security solutions for conducting the evaluation by constructing a list of potential solutions from grey literature and filtering these solutions based on whether they implement some SCM2 functionality, described in the context model (Figure \ref{fig:SCM2_Context}). The grey literature articles were gathered via Google search engine using the search phrase ``Software Supply Chain Security Solutions OR SBOM Tools.'' We include articles that list multiple tools and excluded sponsored search results that link to a specific tool. From these articles, we extracted \textit{34 solutions} potential solutions. The solutions were filtered based on their SCM2 support and ranked according to the number of articles citing them. The top five solutions were utilised for the architecture mapping (Table \ref{tab:evaluation}).

\renewcommand{\arraystretch}{1.0}
\begin{table*}[ht]\footnotesize
\centering
\caption{Architectural Mapping of SSC Security Solutions}
\label{tab:evaluation}
\vspace{-2.00mm}
\resizebox{0.9\linewidth}{!}
{\begin{tabular}{c c c c c c}
\toprule
\textbf{Container} & \textbf{Scribe} & \textbf{Chainguard} & \textbf{Anchore} & \textbf{Snyk} & \textbf{FOSSA} \\
\midrule
\textbf{SSC Metadata Editor}  & View/ Review and Merge (GUI) & View (GUI) & View (GUI) & View (API, CLI) & View (GUI) \\

\textbf{Metadata Generator} & 
SBOM from Container images and Source Code (CycloneDX) & SBOM from Container Images 
  & SBOM from Container Images & SBOM from Container images & SBOM from source code \\
& VeX related to SBOM & (SPDX and CycloneDX) & and Source Code & (SPDX and CycloneDX) & (SPDX and CycloneDX)   \\
& SLSA Provenance & Signed Commits & (SPDX and CycloneDX) & &\\

\textbf{Pipeline Observer}  & \checkmark & \checkmark & - & - & - \\

\textbf{Metadata Signer} & Utilizes Sigstore tools which support in-toto-Attestations & Utilizes Sigstore tools which support 
 & Utilizes Syft SBOM tool which  & - & - \\
 & & in-toto-Attestations & supports in-toto-Attestations & & \\

\textbf{Metadata Publisher} & 
Pub/Sub method & 
Through invites & - & - & 
Shareable links \\

\textbf{Transparency Service} & sigstore Rekor & sigstore Rekor & - & - & - \\

\textbf{Metadata Access Control}  & 
Pub/Sub method &
IAM-based access & - & - & - \\

\textbf{Metadata Repository} & \checkmark
 & \checkmark
  & \checkmark
 & - & \checkmark \\
& Cloud Storage, OCI Registry, Local directory & OCI Regsitry & PostgreSQL & & \\

\textbf{Registry} & Rekor Transparency Log  & Rekor Transparency Log  & - & - & - \\

\textbf{Metadata Access Service} & \checkmark & \checkmark & \checkmark & - & \checkmark \\

\textbf{SSC Metadata Validator} & - & - & - & - & - \\

\textbf{SSC Metadata Verifier} & \checkmark  & \checkmark & - & - & - \\

\textbf{SSC Metadata Resolver} & \checkmark  & \checkmark  & \checkmark  & \checkmark  & \checkmark  \\

\textbf{Deployment} & - & - & 
Distributed & - & Monolithic \\

\textbf{Governance } & Centralised & Centralised & Centralised & Centralised & Centralised \\
\bottomrule

\end{tabular}
}
\end{table*}

\subsection{Architecture Mapping}

\textbf{Scribe Platform:} Scribe\footnote{\url{https://scribesecurity.com/scribe-platform/}} helps software producers and consumers manage risks within their supply chains. It helps them generate evidence such as SBOMs to prove the safety and compliance of their software against regulations and standards such as SLSA. The platform implements the pipeline observer by leveraging the \textit{integration with build systems} like GitHub Actions and Jenkins. Scribe also supports \textit{generating SSC metadata from container images}, such as by leveraging the SBOM generation feature of Docker's BuildKit. Based on inputs from build systems and container image analysis, Scribe can \textit{generate SBOM in CycloneDX format and SLSA-compliant provenance statements}. It also \textit{supports the generation of VeX} to enrich the generated SBOMs. Regarding signing, Scribe supports in-toto attestation by leveraging the \texttt{cosign} tool from the \texttt{sigstore} toolset. Scribe also supports SSC metadata countersigning by leveraging \texttt{Rekor} from the \texttt{sigstore} toolset to implement both the registry and transparency service. Regarding SSC metadata storage and distribution, Scribe supports storing SSC metadata in a cloud storage service or a locally deployed OCI-based container registry. A Pub/Sub protocol implements the advertisement and access control of SSC metadata. On the consumer side, Scribe supports verification of the software attestation by leveraging \texttt{sigstore} tools. 

\vspace{1.5mm}
\noindent\textbf{Chainguard Enforce Platform:} The Chainguard Enforce\footnote{\url{https://www.chainguard.dev/chainguard-enforce}} is an SSC security solution focusing on container workloads. It continuously inspects the SSC metadata of container images within a Kubernetes cluster to evaluate and enforce user-defined policies. Whilst Chainguard Enforce focuses on the consumption of SBOM for policy enforcement, it can also \textit{generate SBOM in SPDX and CycloneDX format} for container images that lack an associated SBOM by leveraging Syft\footnote{\url{https://github.com/anchore/syft}}, an open-source SCA tool for container images. Chainguard Enforce does not extract SSC metadata fragments from build systems. Thus, it \textit{does not include a pipeline observer}. Chainguard Enforce utilises \textit{an OCI-based container registry as the repository} for both containers and the associated SBOM. Producers advertise the existence of SBOM by sending invitations with invite codes. Access is granted via an \textit{Identify and Access Management (IAM) model}. Similar to the Scribe platform, Chainguard Enforce utilises \texttt{cosign} and \texttt{rekor} from the \texttt{sigstore} suite for implementing the signer, registry, and transparency service. \texttt{cosign} is also used to verify the signature of the container images.

\vspace{1.5mm}
\noindent\textbf{Anchore Platform:} Anchore\footnote{\url{https://anchore.com/opensource/}} is an SCM2 focusing on generating and consuming SBOM of container images. Anchore revolves around two open-source tools: Syft for generating SBOM from container images and Grype for consuming SBOM for vulnerability scanning. Anchore does not feature a pipeline observer because it does not capture information about tasks in the build pipeline but triggers the generation of SBOMs and vulnerability analysis upon commits. Anchore utilises PostgreSQL as the data store and supports on-prem databases or integration with external services like Amazon RDS. We could not identify support for SSC metadata signing, notarisation, and signing. 

\vspace{1.5mm}
\noindent\textbf{Snyk Platform:} Snyk\footnote{\url{https://snyk.io/}} is a developer security platform that provides vulnerability detection across the software life cycle. The platform has been extended to support SBOM generation for external consumption. Snyk can \textit{generate SBOM in SPDX and CycloneDX} format. While the Snyk platform has a GUI, it mainly supports viewing vulnerability reports, whereas viewing the generated SBOMs is carried out by accessing a REST API through the curl command line tool or using the Snyk CLI. We could not identify support for SSC metadata signing, notarisation, and signing. 

\vspace{1.5mm}
\noindent\textbf{FOSSA Platform:} FOSSA\footnote{\url{https://fossa.com/}} is an open-source management platform focusing on providing visibility into the licenses and vulnerability of the open source dependencies of a software inventory. FOSSA supports generating SBOM in SPDX and CycloneDX format. It leverages SCA tools to provide fragments to the SBOM generation process. In the SaaS mode of FOSSA, SBOMs are hosted in cloud storage that serves as a repository. Producers can advertise the generated SBOMs by sharing an access link to the repository with consumers. We could not identify support for SSC metadata signing, notarisation, and signing. 

\subsection{Discussions}

The architecture mappings above demonstrated the ability and usability of the SCM2 RA in modelling the functionality and architecture of SCM2 instances. The mappings also reveal some interesting insights about the state of practice of SCM2 and its divergence from academic literature and the existing frameworks. 

\textit{Containers rather than packages:} Three out of five analysed SSC security solutions generate and consume SSC metadata, particularly SBOM, at the software container level. This focus aligns with the cloud native computing paradigm. Nevertheless, it starkly contrasts with the academic literature (e.g., \cite{Ohm2020}) and our experiences in academia-industry collaborative projects, which tend to focus on software packages distributed via package repositories such as NPM and PyPI. An advantage of a container-centric approach is the ease of SSC metadata sharing. The generated SSC metadata documents can be embedded within container images and thus distributed via the same distribution channels. By embedding the SSC metadata inside container images, one digital signature can be used to secure both the artefact (container image) and the SSC metadata document. Furthermore, if a container image carries its corresponding SSC metadata, the resolution step would be simplified as consumers no longer need to link SSC metadata documents with their corresponding software artefacts. 

\textit{Consumers are producers:} Beside Snyk, the analysed SSC security solutions focus on the consumption of SSC metadata for vulnerability detection and policy enforcement. When SSC metadata is generated, it is usually performed in a reverse engineering manner (e.g., analysing an existing container image to extract information and create an SBOM) by the consumers. In other words, the SCM2 implemented by the analysed SSC security solutions belong to the centralised and single-user variant. They exist as utilities of an organisation rather than a diverse ecosystem of SSC metadata producers, providers, and consumers that jointly produce and consume SSC metadata described in the SCITT architecture \cite{Birkholz2023} and the SBOM network concept \cite{NTIA2021}.

\textit{A focus on SBOM:} All analysed SSC security solutions focus on SBOM written in SPDX and CycloneDX format rather than software provenance statements in either SLSA \cite{SLSA2023} or in-toto \cite{intoto2017} provenance model. Even when attestation was utilised, it was used to wrap around SBOM content, such as by using in-toto SPDX predicates\footnote{\url{https://github.com/in-toto/attestation/blob/main/spec/predicates/spdx.md}}. The centralised and single-user design focusing on the consumption of SSC metadata might have contributed to the focus on SBOM rather than provenance. 

\textit{Attestation and notarisation are not commonplace practices:} Signing and notarisation of SSC metadata are not universally supported. Such a state of practice aligns with the requirement levels specified by SLSA \cite{SLSA2023} and SCVS \cite{OWASP2020}, which put attestation at a less immediate level than SSC metadata generation. When signing and notarisation were performed, the primary implementation was the \texttt{sigstore} software suite.

\begin{figure}[ht]
    \centering
    \includegraphics[width=\linewidth]{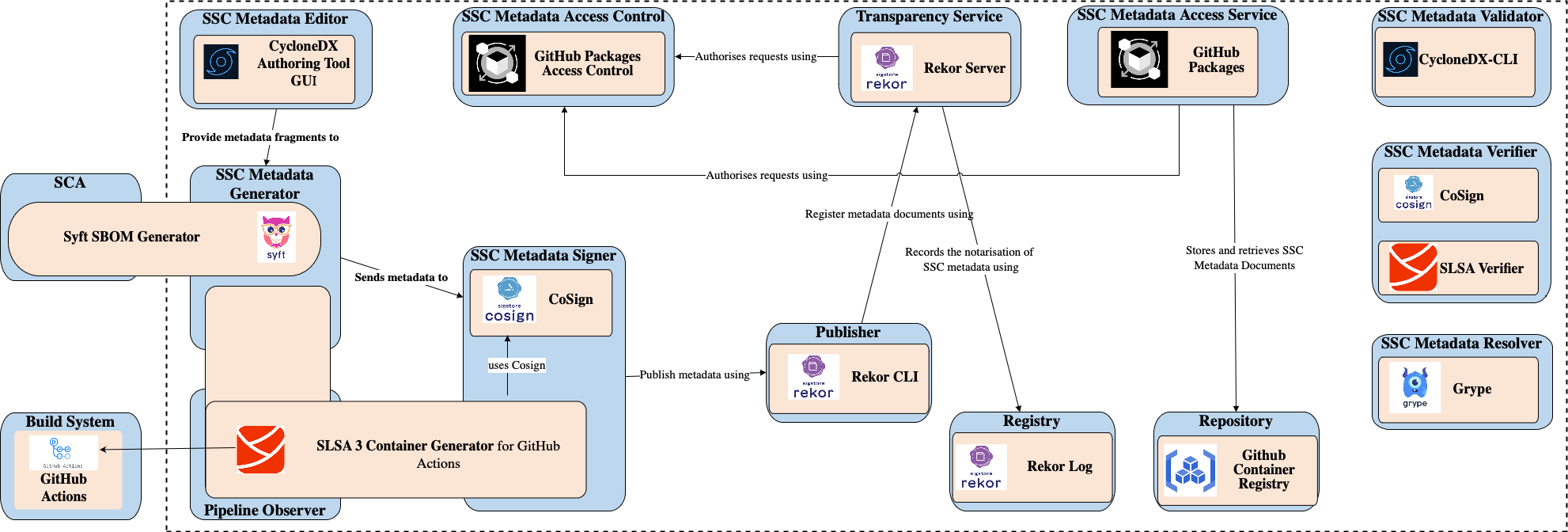}
    \caption{Reference Instantiation of RA}
    \label{fig:reference_instantiation}
    \vspace{-0.5cm}
\end{figure}

\noindent\textbf{Reference Instantiation of SCM2:} Based on the identified open-source tools, we assembled the following reference instantiation of an SCM2. It covers the end-to-end life cycle of all three types of SSC metadata (SBOM, Provenance and Attestation)  from the following SSC layout: software supplier committing the source code to a Version Control System, which is then input into a CI/CD build pipeline to generate a container image as output and publishes it to a Container Image Registry along with SSC metadata related to it. Figure \ref{fig:reference_instantiation} presents the binding of SCM2 containers to open-source tools and technologies.

\section{Conclusions}

SSC metadata can provide machine-readable and authenticated visibility into the SSC of a software asset inventory, aiding software consumers in protecting themselves against SSC attacks. This paper presented an SoK about SSC metadata in the form of an empirically grounded RA for SCM2. Our RA provides practitioners with a comprehensive domain model and an architectural blueprint for communicating SSC metadata requirements with stakeholders and assembling a suitable SCM2. We demonstrated the correctness and utility of the proposed RA by mapping the architecture of five SSC security solutions with SCM2 onto the RA. The architecture mappings also revealed insights about the state of practice, showing a consumer-driven, SBOM-centric approach to SCM2 and SSC metadata. Because SSC security and SCM2 should be a collaborative effort, for our future work, we plan to develop a software suite based on the RA to facilitate a decentralised and multi-user ecosystem around SCM2 where SSC metadata would be generated from the point of origin, authenticated and notarised by a federation of transparency services and discovered and retrieved by authorised consumers via secure channels. Such a decentralised and multi-user SCM2 ecosystem could serve as a ``foundation data layer'' on which further security tools, practices, and assurances can be built.

\section{Data Availability} \label{sec:data}
Online appendix to this paper is available at \textcolor{blue}{\url{https://figshare.com/s/567635a794d7eecf01fe}}. Appendix contains comprehensive details of the SSC security frameworks used for creating the empirical foundation, knowledge graph constructed from the extracted empirical data and high quality images of all figures used in the manuscript.

\section{Acknowledgement} \label{sec:ack}
This work has been supported by the Cyber Security Cooperative Research Centre Limited whose activities are partially funded by the Australian Government’s Cooperative Research Centre Program.

\bibliographystyle{ACM-Reference-Format}
\bibliography{reference.bib}

\end{document}